\documentclass[%
reprint,
,twocolumn,
 showpacs,superscriptaddress,
 amsmath,amssymb,
 aps,
]{revtex4-2}

\usepackage{float}
\usepackage{subcaption}
\usepackage{tikz}
\usetikzlibrary{quantikz}
\usepackage{graphicx}
\usepackage{dcolumn}
\usepackage{bm}
\usepackage{caption}
\newcommand\THEOS{Theory and Simulation of Materials (THEOS), and National Centre for Computational Design and Discovery of Novel Materials (MARVEL), {\'E}cole Polytechnique F{\'e}d{\'e}rale de Lausanne, 1015 Lausanne, Switzerland}
\newcommand\IBMR{IBM Quantum, IBM Research - Zurich, Säumerstrasse 4, 8803 Rüschlikon, Switzerland}
\newcommand\SAP{Dipartimento di Fisica, Universit{\`a} di Roma La Sapienza, Piazzale A. Moro 5, 00185 Roma, Italy}

\begin{document}

\preprint{APS/123-QED}
\title{One-particle Green's functions from the quantum equation of motion algorithm}
\interfootnotelinepenalty=10000
\author{Jacopo Rizzo}
\email[Corresponding author. ]{jrizzo@sissa.it}
\affiliation{\SAP}\affiliation{\THEOS}\affiliation{\IBMR}
\author{Francesco Libbi}
\affiliation{\THEOS}\affiliation{\IBMR}
\author{Francesco Tacchino}
\affiliation{\IBMR}
\author{Pauline J.~Ollitrault}
\affiliation{\IBMR}
\author{Nicola Marzari}
\affiliation{\THEOS}
\author{Ivano Tavernelli}
\affiliation{\IBMR}

\date{\today}

\begin{abstract}
Many-body Green’s functions encode all the properties and excitations of interacting electrons. While these are challenging to be evaluated accurately on a classical computer, recent efforts have been directed towards finding quantum algorithms that may provide a quantum advantage for this task, exploiting architectures that will become available in the near future.
In this work we introduce a novel near-term quantum algorithm for computing one-particle Green’s functions via their Lehmann representation. The method is based on a generalization of the quantum equation of motion algorithm that gives access to the charged excitations of the system.
We demonstrate the validity of the present proposal by computing the Green’s function of a two-site Fermi-Hubbard model on a IBM quantum processor.

\end{abstract}

\maketitle


\section{\label{sec:level1}Introduction}
The exponential scaling of time and memory resources required to accurately simulate an interacting quantum system of many particles poses fundamental limits on what can be achieved by classical algorithms \cite{aaronson2016complexitytheoretic, PhysRevLett.117.080501}. 
This computational burden lies at the heart of many open problems in condensed matter physics, such as the description of high-temperature superconductors \cite{Ding1996} and frustrated magnetic materials \cite{PhysRevLett.113.117201}, for which an efficient and reliable computational treatment is still missing.
In this context, quantum computing has recently emerged as a new promising paradigm~\cite{nielsen_chuang_2010} that can potentially outperform classical counterparts on selected tasks~\cite{Montanaro_2016}, including the simulation of quantum systems~\cite{Feynman1982,Lloyd1996,RevTacchino,PRXQuantum.2.017003}. In the near term, quantum computers may reliably control only a relatively small amount of qubits and perform a limited set of operations on them before significantly loosing coherence \cite{Preskill2018quantumcomputingin}. Nevertheless, these devices are quickly approaching the stage at which it becomes impossible to simulate their behavior classically~\cite{IBMroadmap}. This motivates the search for quantum algorithms that can effectively exploit such limited but potentially useful computational power. To this aim, one should in general design methods which require a relatively low qubit and gate count, but that can still offer a quantum advantage by accessing otherwise classically intractable regimes~\cite{bharti2021noisy, McClean_2016}.

The calculation of static and dynamical properties of many-body quantum systems is in fact one of the most promising research areas in which quantum computers could offer significant advantages. In this context, several near-term methods have been proposed, starting from the well known variational quantum eigensolver (VQE) for calculating the ground state of a given Hamiltonian \cite{peruzzo2014, McClean_2016}. Quite generally, these methods optimize the use of quantum resources by splitting the target problems into parts, distinguishing between those that can be efficiently solved on a classical computer and the ones~\cite{Rossmannek2021embedding,eddins2021doubling} that instead should be delegated to quantum processors.
Promising quantum solutions have also been proposed and implemented for the evaluation of the excitation spectra~\cite{Jones_2019, Higgott_2019, Nakanishi_2019, Parrish_2019, McClean_2017, qeom_IBM}, for variational time evolution~\cite{Yuan_2019} and, more recently, to compute dynamic correlation functions \cite{chiesa_quantum_2019,crippa_simulating_2021,Google2021InfoScrambling,PhysRevResearch.2.033281,chen2021variational}. The latter are of crucial importance for the study of optical, magnetic, and transport properties of many-body quantum systems~\cite{giuliani_vignale_2005}. 
As an alternative to approaches based on the many-electron wavefunction, Green's function techniques can provide solutions for the calculation of interesting physical properties, which in the frequency domain can be directly accessed and validated in photoemission experiments.

In this work, we propose a novel near-term quantum algorithm for computing the Green’s function via its Lehmann representation, following the general setup presented by Endo \textit{et al.}~in Ref.~\cite{PhysRevResearch.2.033281}. 
Our method relies on an original extension of the quantum equation of motion (qEOM) algorithm \cite{qeom_IBM} that allows us to compute the charged excited states of the system. The latter are used to efficiently compute the spectral amplitudes needed to evaluate the Green’s function in the frequency domain. We test our proposed quantum algorithm numerically and experimentally on IBM Quantum devices by computing the Green’s function of a two-site Fermi-Hubbard model.

The rest of this paper is organized as follows. In Sec.~\ref{sec:gf} we fix the notation for the formalism of many-body Green's functions. In Sec.~\ref{sec:gf_qEoM} we describe our proposed method for computing the Green's function on a quantum computer via the qEOM technique. In Sec.~\ref{sec:simulations} we demonstrate the validity of our proposal by performing numerical and hardware simulations. Finally, in Sec.~\ref{sec:conclusions} we conclude by commenting on possible future improvements and research directions.

\section{Many-body Green's functions}\label{sec:gf}
We start by reviewing the formalism of many-body Green's functions at zero temperature for fermionic quantum systems \cite{Fetter2003Quantum}.
We consider a system of $N$ identical fermions defined by its Hamiltonian $\hat{H}$, projected onto a basis of $M$ fermionic modes $\{ | \psi_{\alpha} \rangle \}_{\alpha=1}^{M}$. In this context we may introduce the fermionic creation (destruction) operators $\hat{c}_{\alpha}^{\dagger}$ ($\hat{c}_{\alpha}$) satisfying the usual anti-commuting algebra, and we can define the single-particle \textit{retarded Green's function} (GF) at zero temperature as \cite{giuliani_vignale_2005}
\begin{equation}
G^{R}_{\alpha \beta}(t) = - i \theta(t) \langle \psi_0 | \{ \hat{c}_{\alpha}(t), \hat{c}_{\beta}^{\dagger}(0) \} | \psi_0 \rangle.
\end{equation}
Here $\theta(t)$ is the Heaviside step function, $\hat{c}_{\alpha}(t) = e^{i\hat{H}t} \hat{c}_{\alpha} e^{-i\hat{H}t}$ is the Heisenberg evolution of the operator $\hat{c}_{\alpha}$ and the expectation value is computed over the $N$ particle ground state of the Hamiltonian $| \psi_0 \rangle$. In our definition we take $\hbar = 1$. 
The GF as defined in Eq.~(1) may be expressed in Fourier space by introducing the regularization factor $\lim_{\eta \rightarrow 0^+} e^{- \eta |t|}$ as
\begin{equation}
\tilde{G}^{R}_{\alpha \beta}(\omega) = \int_{-\infty}^{\infty} G^{R}_{\alpha \beta}(t) e^{i \omega t - \eta |t|} dt.
\end{equation}
The real frequency GF then becomes \cite{giuliani_vignale_2005}
\begin{equation}
\begin{aligned}
\tilde{G}^{R}_{\alpha \beta}(\omega) = \langle \psi_0| \hat{c}_{\alpha} \frac{1}{\omega + i \eta + E_{0} - \hat{H}} \hat{c}_{\beta}^{\dagger} | \psi_0 \rangle \\
+ \langle \psi_0| \hat{c}_{\beta}^{\dagger} \frac{1}{\omega + i \eta + \hat{H} - E_{0} }  \hat{c}_{\alpha} | \psi_0 \rangle.
\end{aligned}
\end{equation}
If we insert in Eq.~(3) the representation of the identity $I = \sum_n |E_n \rangle \langle E_n |$, where $|E_n\rangle$ are the (orthonormal) eigenstates of the system, Eq.~(3) can be written as
\begin{equation}
\begin{aligned}
\tilde{G}^{R}_{\alpha \beta}(\omega) = \sum_n \biggl(  \frac{\langle \psi_0| \hat{c}_{\alpha} | E_n^{N+1} \rangle \langle E_n^{N+1} | \hat{c}_{\beta}^{\dagger} | \psi_0 \rangle }{\omega + i \eta + E_{0}^N - E_n^{N+1}}  
\\
+  \frac{ \langle \psi_0| \hat{c}_{\beta}^{\dagger} | E_n^{N-1} \rangle \langle E_n^{N-1} | \hat{c}_{\alpha} | \psi_0 \rangle}{\omega + i \eta + E_n^{N-1} - E_{0}^N } \biggl) ,
\end{aligned}
\end{equation}
where $E_n^{N+1} (E_n^{N-1}) $ are the eigenstates of the system with $N+1$ $(N-1)$ particles, also referred as the particle (hole) spectrum.
Eq.~(4) constitutes the \textit{Lehmann representation} of the retarded GF.
In what follows, for simplicity, we restrict our attention to the diagonal terms of the GF
\begin{equation}
\begin{aligned}
\tilde{G}^{R}_{\alpha \alpha}(\omega) = \tilde{G}^{R}_{\alpha}(\omega) = \sum_{n} \biggl(  \frac{|\langle E_n^{N+1} | \hat{c}_{\alpha}^{\dagger} | \psi_0 \rangle|^2}{\omega + E_0^{N} - E_n^{N+1} + i \eta} \\ 
+ \frac{|\langle E_n^{N-1} | \hat{c}_{\alpha} | \psi_0 \rangle|^2}{\omega - E_0^{N} + E_n^{N-1} + i \eta}   \biggl).
\end{aligned}
\end{equation}
In fact, the generalization to Eq.~(4) of the method we propose can be easily accomplished.
The single-particle GF in the real frequency domain is an interesting quantity, since its imaginary part, the \textit{spectral function}, is directly accessible via photoemission spectroscopy \cite{giuliani_vignale_2005}
\begin{equation}
A_{\alpha}(\omega) = - \pi^{-1} \operatorname{Im}(\tilde{G}^{R}_{\alpha}(\omega)).
\end{equation}
In general, the single-particle GF characterizes the dynamics of quasiparticles in correlated systems \cite{Fetter2003Quantum}. Furthermore, other types of GF can be used to characterize the optical, magnetic and transport properties of a system in the framework of \textit{linear response theory} \cite{giuliani_vignale_2005}, where the effect of an external perturbation $\hat{B} V(t)$ on an observable $\hat{A}$ can be expressed as
\begin{equation}
\mathcal{A}(t) = \mathcal{A}(t_0) + \int_{-\infty}^{\infty} \chi_{AB}(t-t') V(t') dt',
\end{equation}
where
\begin{equation}
\chi_{AB}(t) = -i \theta(t) \langle \psi_0 | [\hat{A}(t), \hat{B} ]| \psi_0 \rangle,
\end{equation}
Where $[\hat{A},\hat{B}]=\hat{A}\hat{B}-\hat{B}\hat{A}$, $\hat{A}(t) = e^{i\hat{H}t} \hat{A} e^{-i\hat{H}t}$ and $V(t)$ is the time-dependent envelope of the external perturbation. Eq.~(8) can be understood as a generalization of Eq.~(3). 
The quantum algorithm proposed in this work (see Sec. III) evaluates the GF in the frequency domain via its Lehmann representation Eq.~(5) by first evaluating the excitation spectra of the system and then computing the corresponding spectroscopic amplitudes.

\section{Calculation of the Green's Function via the quantum equation of motion algorithm}\label{sec:gf_qEoM}
The classical equation of motion (EOM) method was initially introduced by Rowe in Ref.~\cite{rowe} as a way to obtain excitation energies and excited states of a given Hamiltonian $\hat{H}$. In the EOM approach we search for an approximation to the excitation operator $\hat{O}_n^{\dagger} = |n \rangle \langle 0 |$,
which generates the $n$-th excited state $|n \rangle$ of the Hamiltonian starting from a given generic state $|\tilde{0} \rangle$ with non-zero overlap with the true ground state $|0 \rangle$.
The excitation operator satisfies
\begin{equation}
[\hat{H},\hat{O}_n^{\dagger}] |\tilde{0} \rangle = E_{0n} \hat{O}_n^{\dagger} |\tilde{0} \rangle,
\end{equation}
where $E_{0n} = E_n - E_0$ are the exact excitation energies. If we take $|\tilde{0} \rangle = | 0 \rangle$, where $|0\rangle$ is the true ground state of the system, the annihilation condition $ \hat{O}_n |\tilde{0} \rangle = 0 $ is satisfied,
and by operating from the left-hand side of Eq.~(9) with $\hat{O}_n^{\dagger} | 0 \rangle$ we may write
\begin{equation}
 E_{0n} = \frac{\langle 0 | [ \hat{O}_n , [\hat{H},\hat{O}_n^{\dagger}]] | 0 \rangle}{\langle 0 |[\hat{O}_n, \hat{O}_n^{\dagger}]  | 0 \rangle}.
\end{equation}
This equation holds if the excitation operator is exact and if $|0\rangle$ is the true ground state. Using again the annihilation condition, we can equivalently write
\begin{equation}
 E_{0n} = \frac{\langle 0 | [ \hat{O}_n , \hat{H},\hat{O}_n^{\dagger}] | 0 \rangle}{\langle 0 |[\hat{O}_n, \hat{O}_n^{\dagger}]  | 0 \rangle},
\end{equation}
where we have introduced the \textit{double commutator} $[\hat{A},\hat{B},\hat{C}] = \frac{1}{2}([[\hat{A},\hat{B}],\hat{C}] + 
[\hat{A},[\hat{B},\hat{C}]])$ and used the fact that $\langle  0| [ \hat{O}_n , [\hat{H},\hat{O}_n^{\dagger}]] | 0 \rangle = \langle 0 | [[\hat{O}_n , \hat{H}],\hat{O}_n^{\dagger}] | 0 \rangle$ if the annihilation condition is satisfied.
This second formulation has the advantage of guaranteeing real-valued energy differences, since the operator $[ \hat{O}_n , \hat{H},\hat{O}_n^{\dagger}]$ in Eq.~(11) is Hermitian.
We can find approximate solutions to Eq.~(11) by considering it as functional of the excitation operator $\hat{O}_n$ that should be minimized. In fact, by taking a variation $\delta \hat{O}_n$ and imposing the stationary condition $\delta E_{0n} = 0$ we get again Eq.~(9), namely the functional of Eq.~(11) is stationary when evaluated at the exact excitation operators. 
We can then approximate the excitation operators by linearly expanding them in a basis of elementary excitations.
For example, if one wants to target the charged excitations of the system one may consider the ansatz
\begin{equation}
\hat{O}_n^{\dagger} = \sum_{\alpha} \sum_{\mu_{\alpha}} X_{\mu_{\alpha}}^{(\alpha)}(n) \hat{E}_{\mu_{\alpha}}^{(\alpha)},
\end{equation}
where $\mu_{\alpha}$ is a collective index of the single-particle orbitals entering in the charged excitation operator $\hat{E}_{\mu_{\alpha}}^{(\alpha)}$ of degree $\alpha$ with coefficient $X_{\mu_{\alpha}}^{(\alpha)}(n)$.
For example, $\hat{E}_{\mu_{\alpha}}^{(\alpha)}$ may be restricted to $\alpha = 0,1,2$ with $\hat{E}_{\mu_{0}}^{(0)} =  \hat{c}^{\dagger}_m$, $\hat{E}_{\mu_{1}}^{(1)} =  \hat{c}^{\dagger}_m \hat{c}^{\dagger}_n \hat{c}_i$ and $\hat{E}_{\mu_{2}}^{(2)} =  \hat{c}^{\dagger}_m \hat{c}^{\dagger}_n \hat{c}^{\dagger}_l \hat{c}_i \hat{c}_j$.
By imposing the stationary condition on Eq.~(11) after inserting  Eq.~(12) we obtain the generalized eigenvalue problem (GEP) equation
\begin{equation}
\textbf{A} \textbf{X}_n = E_{0n} \textbf{B} \textbf{X}_n,
\end{equation}
where
\begin{equation}
A_{\mu_{\alpha}, \nu_{\beta}} = \langle 0|[(\hat{E}_{\mu_{\alpha}}^{(\alpha)})^{\dagger},\hat{H},\hat{E}_{\nu_{\beta}}^{(\beta)} ]   |0\rangle,
\end{equation}
\begin{equation}
B_{\mu_{\alpha}, \nu_{\beta}} = \langle 0|[(\hat{E}_{\mu_{\alpha}}^{(\alpha)})^{\dagger},\hat{E}_{\nu_{\beta}}^{(\beta)} ]   |0\rangle.
\end{equation}
In this case, if the ground state has $N$ particles one targets the sector with $N+1$ particles. We note that, since the functional of Eq.~(11) can be intended both for $\hat{O}_n^{\dagger}$ and $\hat{O}_n$, one also gets the excitation energies corresponding to the $N-1$ particles sector. This is particularly useful for the calculation of the GF in the form of Eq.~(5).

A quantum version of the EOM algorithm, the quantum equation of motion method (qEOM), was introduced by Ollitrault \textit{et al.}~in Ref.~\cite{qeom_IBM}, where the ground state of the system is prepared via the VQE and the matrix elements of the EOM entering in Eq.~(13) are then estimated directly on a quantum computer. This last step can be efficiently done by using for example the operator averaging technique \cite{McClean_2016} for evaluating expectation values on a state prepared on quantum hardware.
The qEOM method displays two significant advantages with respect to the classical EOM. First, the matrix elements entering in the secular problem, see Eq.~(13), can be efficiently measured on a quantum computer under a suitable qubit encoding, while for large systems it is generally unfeasible to evaluate them with classical methods. Second, the qEOM can also rely, in principle, on better representations of the ground state itself. Indeed, while for the EOM one should only consider a ground state which can be easily manipulated classically, the qEOM has access to a quantum register which can host highly entangled quantum states, prepared e.g.~via the VQE. Both these observations are crucial in establishing a possible quantum advantage brought by the qEOM method with respect to its classical counterpart.

In Ref.~\cite{qeom_IBM} the qEOM algorithm was experimentally tested for the evaluation of neutral excitations of small molecules, and was shown to be robust to noise and more accurate than quantum subspace expansion \cite{McClean_2020} in numerical tests.
In this work, we propose a generalized version of the qEOM algorithm which specifically targets charged excitations. This will be used to compute the spectroscopic elements entering in the GF, using an ansatz of the form of Eq.~(12), which can be expressed as
\begin{equation}
\frac{\langle 0 | \hat{O}_n \hat{c}_{\alpha}^{\dagger}|0 \rangle}{\sqrt{\langle 0 | \hat{O}_n \hat{O}_n^{\dagger}|0 \rangle}} = \frac{\langle n | \hat{c}_{\alpha}^{\dagger} |0 \rangle}{ \sqrt{\langle n | n \rangle}},
\end{equation}
where $|0 \rangle$ is the approximate and normalized ground state of the system, obtained via the VQE in this case. The normalization factor is included to ensure that the excited states obtained via qEOM are also normalized, as it is implied by the Lehmann representation. The formula above shows that, to estimate the GF, one should first obtain an approximation to the $N$ particle ground state via the VQE method, and then apply the qEOM algorithm on a set of elementary excitation operators. Finally, one can efficiently access the spectroscopic elements entering in the GF by computing the expectation value over the ground state as written in Eq.~(16).
Compared to other quantum algorithms for computing excitations of the system such as subspace search \cite{Nakanishi_2019} or overlap based methods \cite{Higgott_2019}, the qEOM approach has the advantage of allowing the evaluation of the spectroscopic amplitudes with no extra quantum resources compared to the VQE. This feature is shared, e.g., with the quantum subspace expansion method, although the latter was shown to be less suited than qEOM for molecular simulations~\cite{qeom_IBM}. On the flip side, qEOM requires a rather large of number of measurements to reconstruct the relevant matrices. This, in the true spirit of near-term quantum solutions, shifts part of the computational burden to classical resources, and could be partially alleviated via optimized measurement protocols.
In the next section, we discuss the results of the experimental and numerical implementation of our algorithm applied to the two site Fermi-Hubbard model.

\section{Numerical and hardware simulations}\label{sec:simulations}
We now consider the Fermi-Hubbard model with $N=2$ sites, in the basis
$\{ |1,\uparrow \rangle, |2,\uparrow \rangle, |1,\downarrow \rangle, |2,\downarrow \rangle \} $, with the on-site repulsive interaction $U$ and the hopping $t$
\begin{equation}
\hat{H} = -t \sum_{\sigma = \uparrow,\downarrow} \biggl( \hat{c}_{1,\sigma}^{\dagger} \hat{c}_{2,\sigma} + 
\hat{c}_{2,\sigma}^{\dagger} \hat{c}_{1,\sigma} \biggl) \ + \ U \sum_{i=1}^2 \hat{n}_{i,\uparrow} \hat{n}_{i,\downarrow}.
\end{equation}
Here $\hat{n}_{i,\sigma} = \hat{c}_{i,\sigma}^{\dagger} \hat{c}_{i,\sigma} $ is the occupation number operator and $U,t > 0 $. In our simulations we always take $U=3$ and $t=1$. 
We start by computing the ground state of Hamiltonian in Eq.~(17) via the VQE procedure. To this aim, we map the Hamiltonian from the fermionic Fock space to the Hilbert space of qubits using the Jordan-Wigner (JW) transformation \cite{jordanwigner}. After this step, the Hamiltonian can be expressed as a linear combination of $11$ Pauli strings. These Pauli strings can then be divided in $3$ groups where the Pauli strings belonging to the same group commute with each other. We exploit this grouping when measuring the VQE cost function by applying the standard operator averaging technique and measuring simultaneously the Pauli strings entering in the same group. This allows us to run only $3$ quantum circuits to estimate the expectation value of the energy for each VQE optimization step.

The encoding of the two-site Fermi-Hubbard model requires $4$ qubits. We choose the heuristic ansatz state~\cite{Barkoutsos_2018} shown in Fig. 1, with $4$ parameters to be optimized. Here, the $A$ entangling blocks are of the form
\begin{equation}
A(\theta,\phi) = \begin{pmatrix} 1 & 0 & 0 & 0 \\ 0 & \cos{\theta} & e^{i\phi} \sin{\theta} & 0 \\ 0 & e^{-i\phi}\sin{\theta} & -\cos{\theta}  & 0
\\ 0 & 0 & 0 & 1 \end{pmatrix},
\end{equation}
and can be expressed using the circuit
\begin{center}
\begin{quantikz}
\qw & \targ{} & \qw & \ctrl{1} & \qw & \targ{} & \qw &\\
 \qw & \ctrl{-1} & \gate{R(\theta,\phi)} & \targ{} & \gate{R(\theta,\phi)^{\dagger}} & \ctrl{-1} & \qw &\\
\end{quantikz}
\end{center}
where $R(\theta,\phi) = R_{y}(\theta + \pi/2) R_{z}(\phi + \pi)$. We notice that this ansatz preserves both the particle number and the spin projection $S_z$. To enforce specific values for these observables, the two $X$ gates at the beginning of the circuit initialize the state with the correct spin projection ($S_z = 0$) and number of particles $N = 2$.
The circuit consists of 8 CNOT gates (6 entering in the $A$ operators, $2$ external). The actual number of CNOT gates then amounts to 11 on \textit{ibm\_cairo}, due to the limited connectivity of the hardware. We show a scheme of the quantum processor in Fig. 2, highlighting the $4$ qubits we use to implement our algorithm.
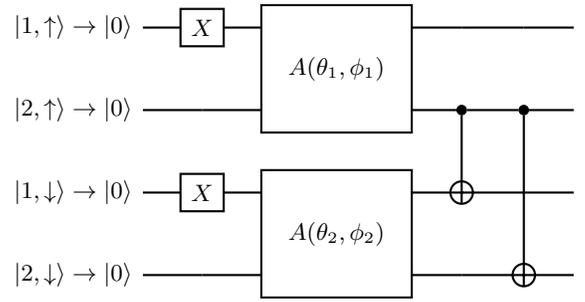
\begin{figure}[h]
\begin{center}
\begin{quantikz}
\lstick{$ \ket{1,\uparrow} \rightarrow \ket{0}$} & \gate{X} & \gate[wires=2][2cm]{A(\theta_1,\phi_1)} & \qw & \qw & \qw \\
\lstick{$ \ket{2,\uparrow} \rightarrow \ket{0}$} & \qw & & \ctrl{1} & \ctrl{2} & \qw\\
\lstick{$ \ket{1,\downarrow} \rightarrow \ket{0}$} & \gate{X} & \gate[wires=2][2cm]{A(\theta_2,\phi_2)} & \targ{} & \qw & \qw\\
\lstick{$ \ket{2,\downarrow} \rightarrow \ket{0}$}  & \qw & & \qw & \targ{} & \qw\\
\end{quantikz}
\end{center}
\caption{Ansatz circuit used to represent the ground state of the two-site Fermi-Hubbard model with two electrons in the VQE scheme. The particle-number conserving $A$ gates are those defined in Eq.~(18). The $X$ gates set the correct values for the number of particles and the spin projection $S_z$. The overall circuit then conserves these quantities.}
\end{figure}
\begin{figure}[h] 
\centering
\includegraphics[width = \columnwidth]{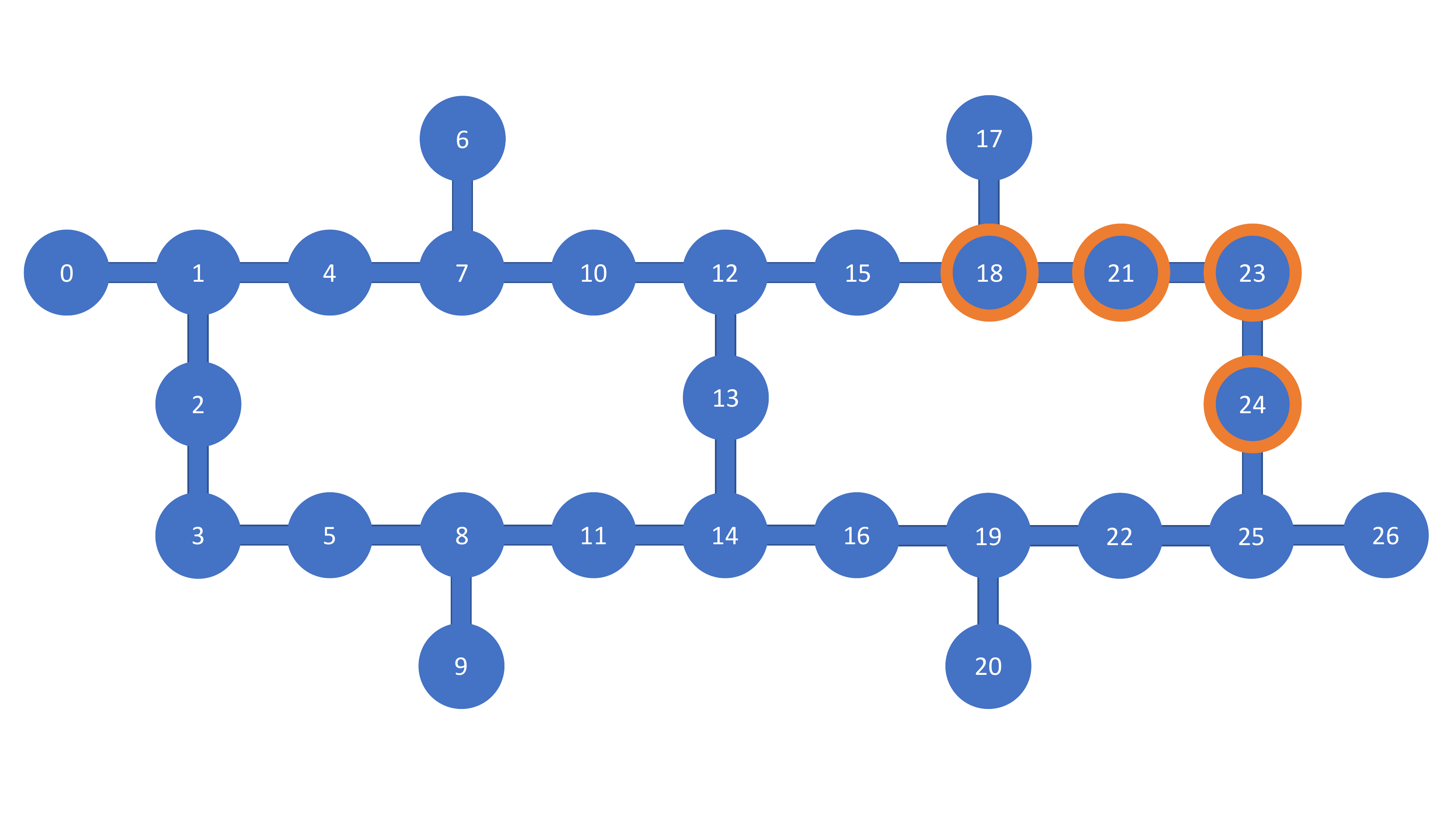}
\centering
\caption{Picture of the topology of the \textit{ibm\_cairo} 27 qubit device. We highlight in orange the qubits used to implement our algorithm.}
\end{figure}
\\ The two $A$ gates perform a rotation in the subspace with fixed number of particles and $S_z$ for each of the two qubit sets having the same spin (up or down). Moreover, the two CNOT gates generate enough entanglement between the two sets to represent the correct ground state for the Fermi-Hubbard dimer, while preserving particle number and spin projection. 
When performing the VQE, single-qubit rotations should be attached to the ansatz circuit in Fig.~1, before measuring in the computational basis to estimate the VQE cost function at each optimization step.

We execute four types of simulations for the VQE and for the overall algorithm. We start performing statevector and qasm simulations in Qiskit~\cite{Qiskit}. We then equip the qasm simulator with a noise model (via the qiskit \texttt{NoiseModel.from\_backend} functionality) taking into account the noise levels of the actual hardware. Finally we test our algorithm on the \textit{ibm\_cairo} quantum computer.
For what concerns the VQE energy minimization procedure, we use the COBYLA
\begin{figure*} 
\begin{subfigure}{\columnwidth}
  \includegraphics[width=\columnwidth]{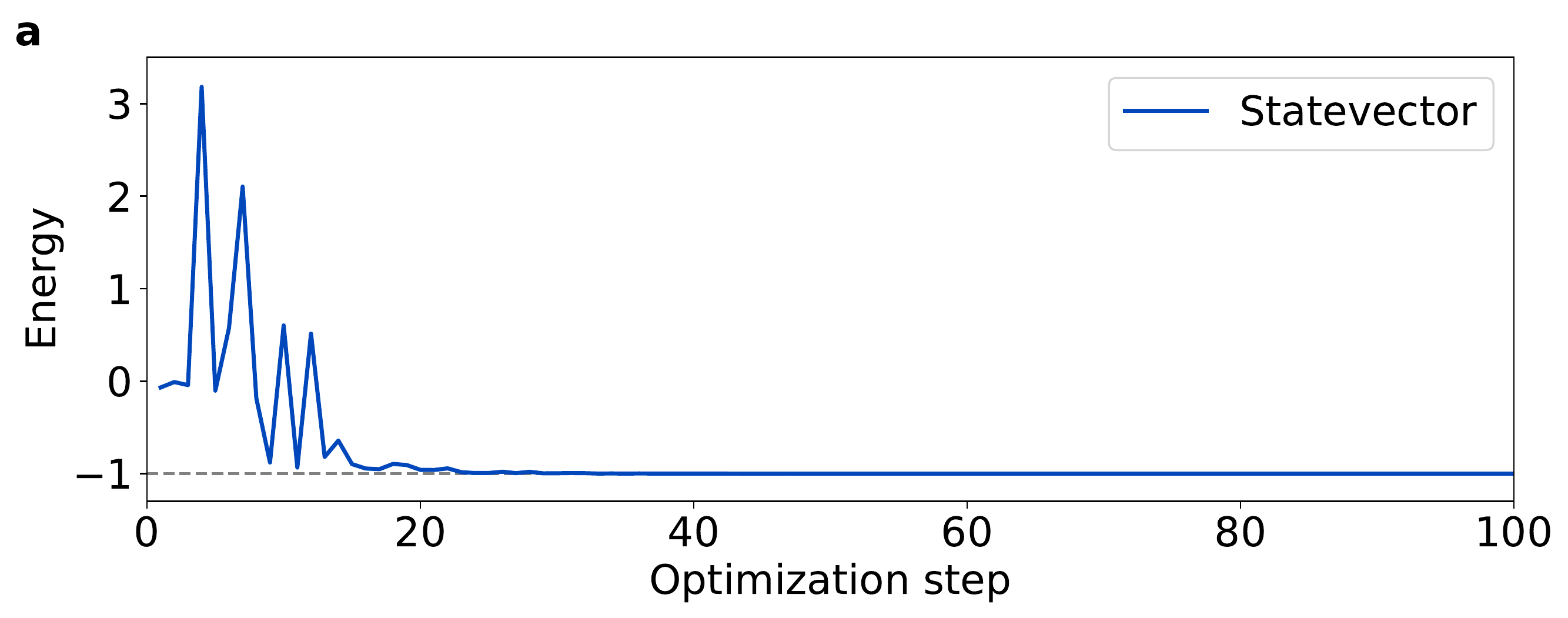}  
\end{subfigure}
\begin{subfigure}{\columnwidth}
  \includegraphics[width=\columnwidth]{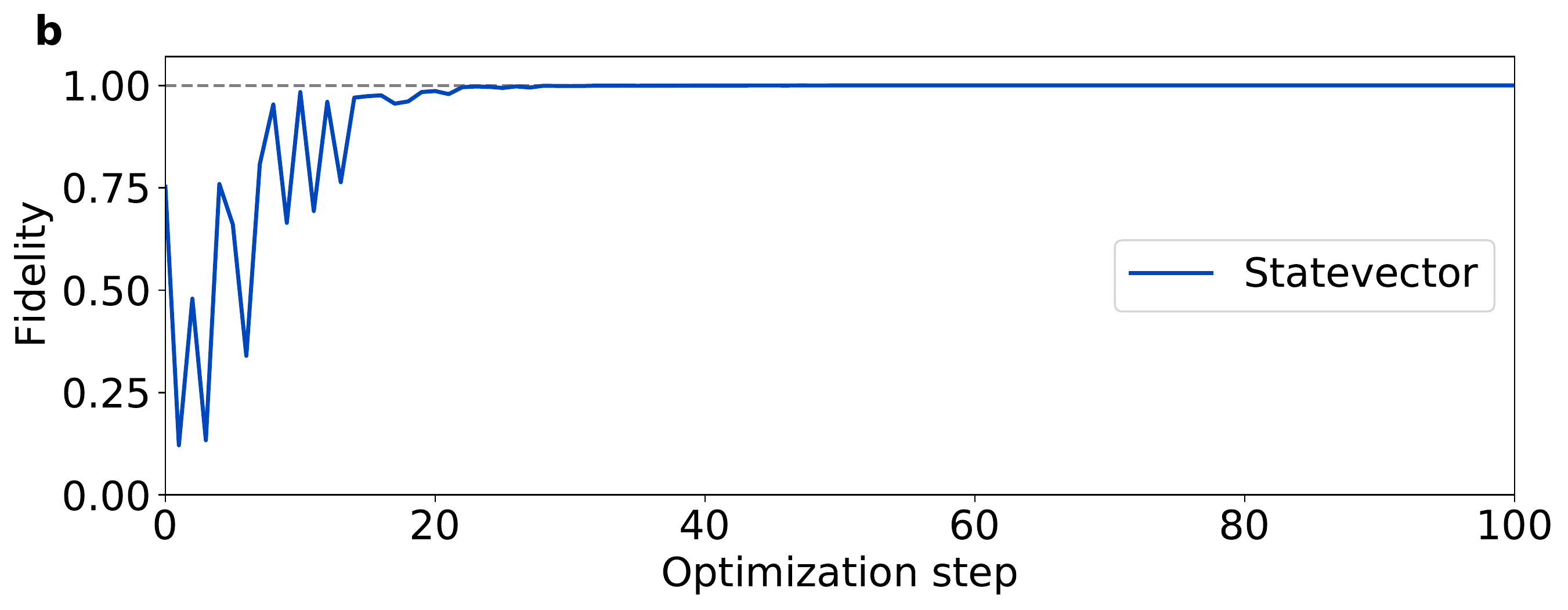}  
\end{subfigure}
\begin{subfigure}{\columnwidth}
  \includegraphics[width=\columnwidth]{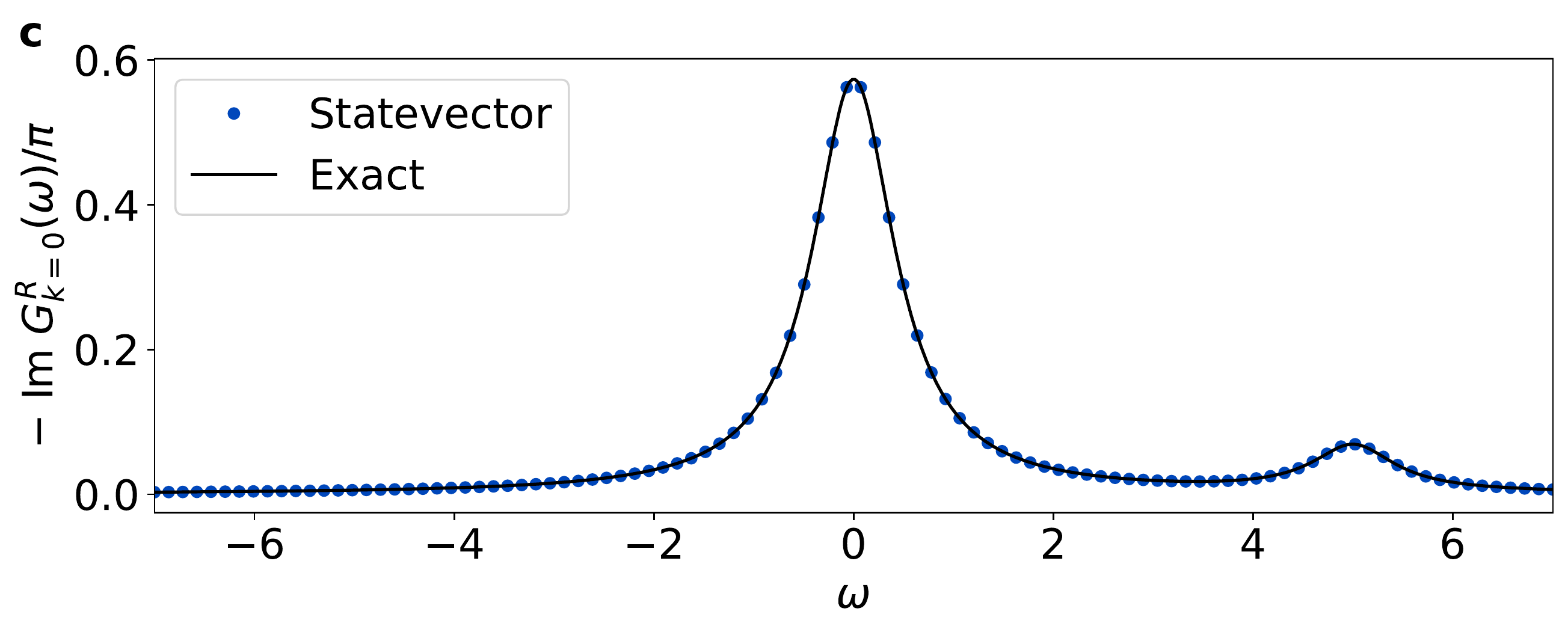}  
\end{subfigure}
\begin{subfigure}{\columnwidth}
  \includegraphics[width=\columnwidth]{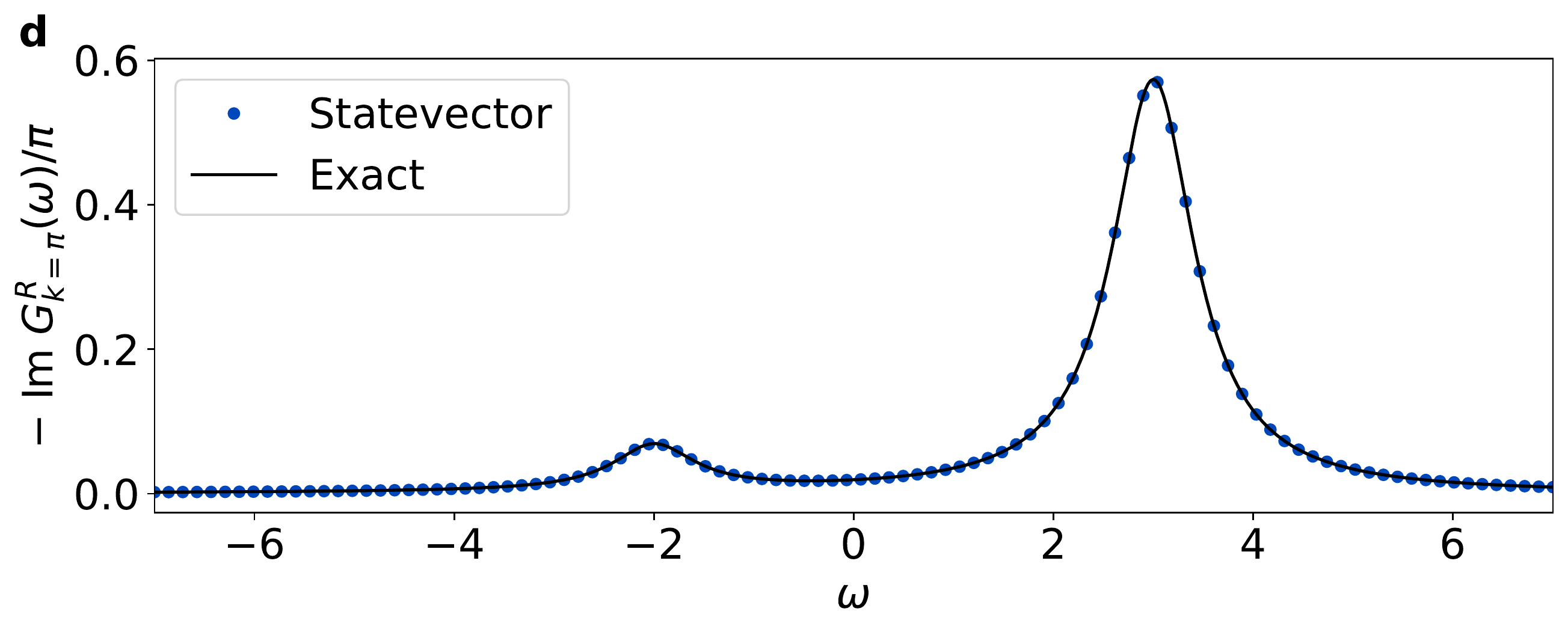}  
  \end{subfigure}
\caption{(\textbf{a}) Statevector estimate of the energy at each optimization step of the VQE procedure. The dashed line is the exact ground state energy. (\textbf{b}) Fidelity between the quantum state represented by the circuit at each time step and the exact ground state. The dashed line is the optimal fidelity. (\textbf{c})-(\textbf{d}) Statevector simulation (and exact result) of the imaginary part of the retarded Green's function for $k=0$ (\textbf{c}) and $k=\pi$ (\textbf{d}) with $\eta=0.5$ via the qEOM algorithm.}
\end{figure*}
\begin{figure*}
\begin{subfigure}{\columnwidth}
  \includegraphics[width=\columnwidth]{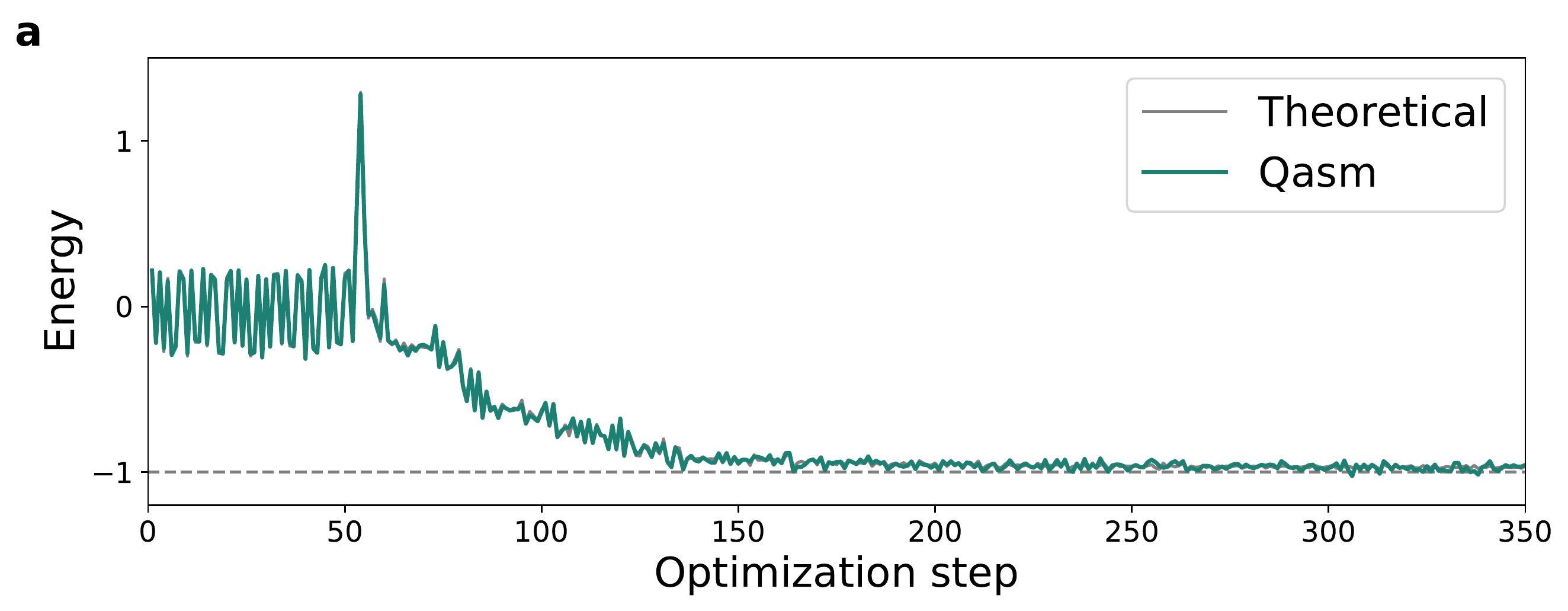}  
\end{subfigure}
\begin{subfigure}{\columnwidth}
  \includegraphics[width=\columnwidth]{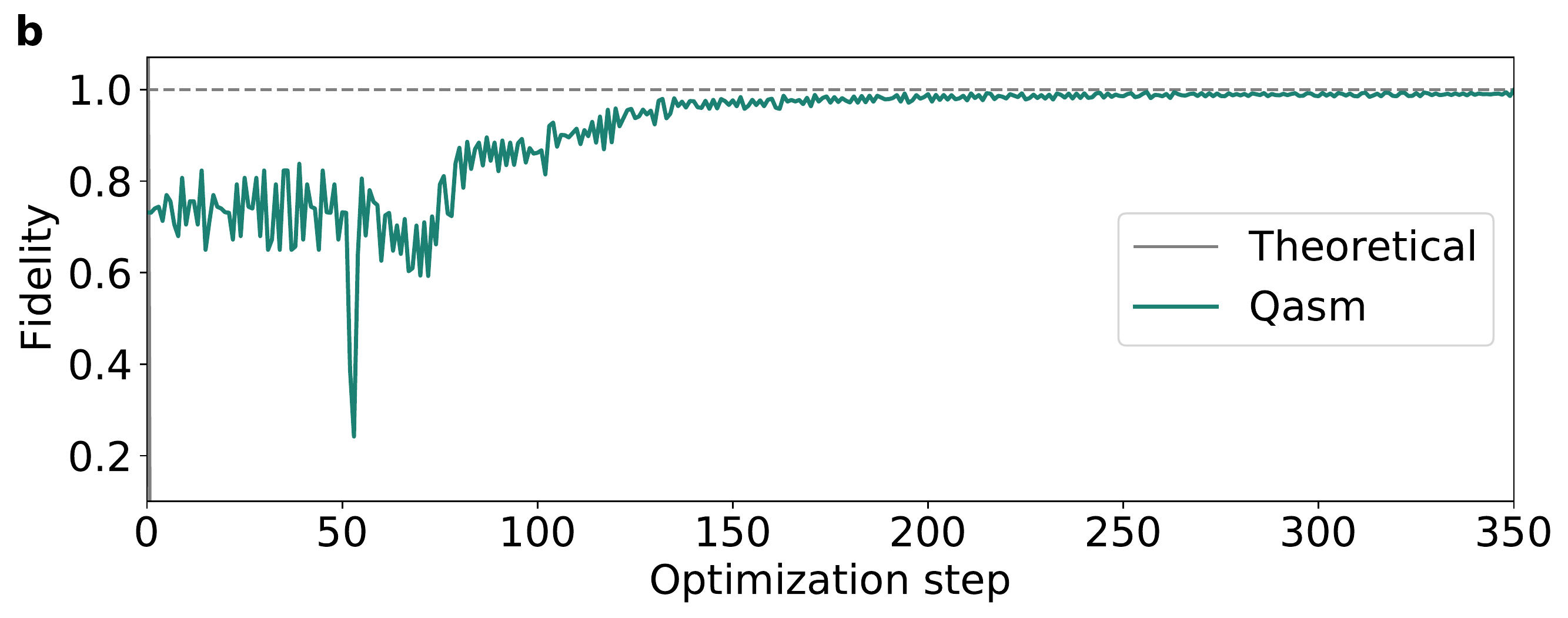}  
\end{subfigure}
\begin{subfigure}{\columnwidth}
  \includegraphics[width=\columnwidth]{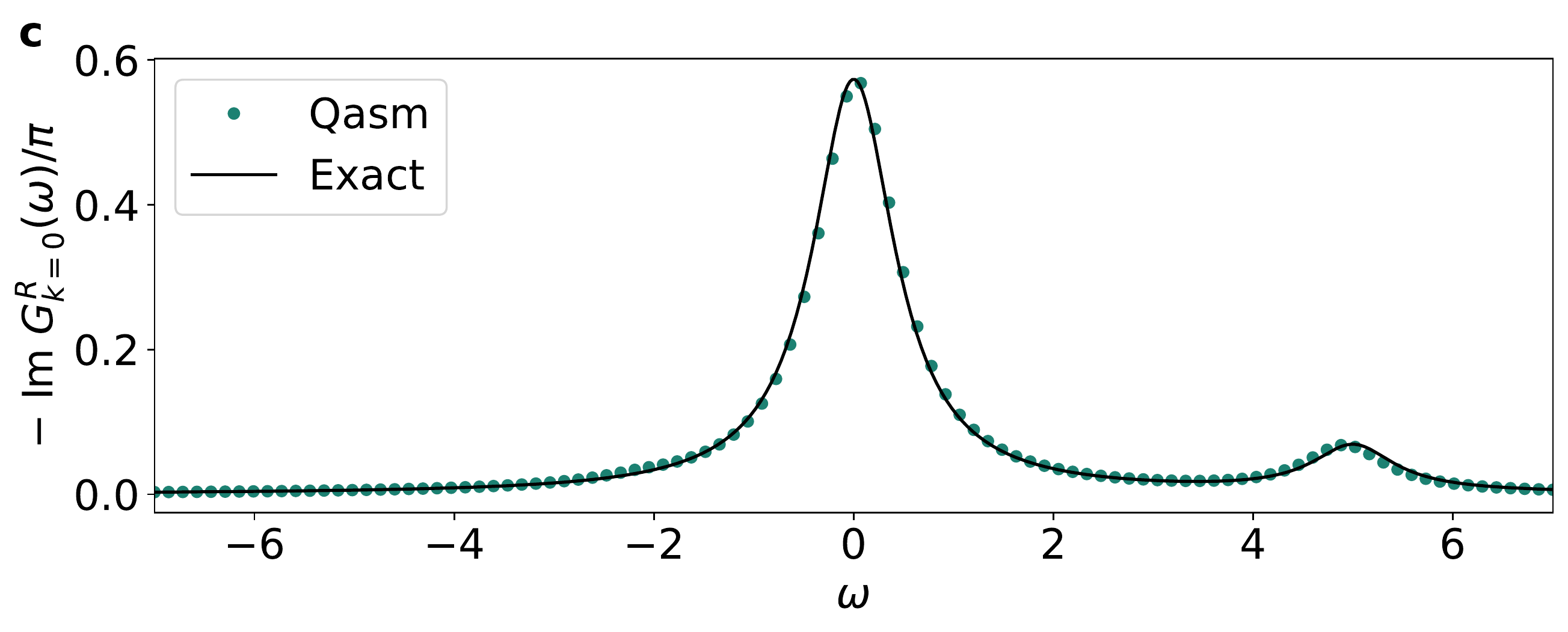}  
\end{subfigure}
\begin{subfigure}{\columnwidth}
  \includegraphics[width=\columnwidth]{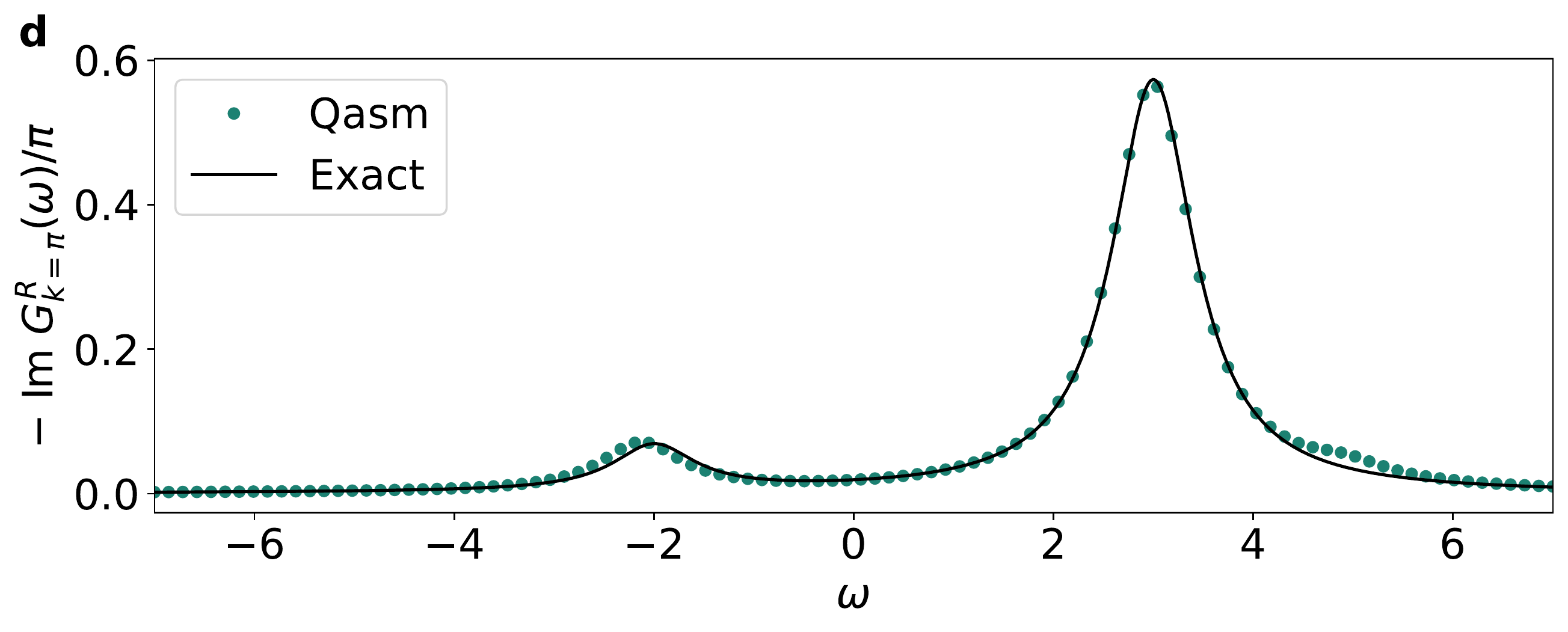}  
  \end{subfigure}
\caption{(\textbf{a}) Theoretical and qasm estimate of the energy at each optimization step of the VQE procedure. The dashed line is the exact ground state energy. (\textbf{b}) Theoretical and 2asm fidelity (obtained via QST) between the quantum state represented by the circuit at each time step and the exact ground state. The dashed line is the optimal fidelity. (\textbf{c})-(\textbf{d}) Qasm simulation (and exact result) of the imaginary part of the retarded Green's function for $k=0$ (\textbf{c}) and $k=\pi$ (\textbf{d}) with $\eta=0.5$ via the qEOM algorithm.}
\end{figure*}
\cite{Powell1994} classical optimizer for statevector simulations, while qasm, noisy qasm and hardware simulations are performed employing the SPSA algorithm \cite{119632}, which performs better in the presence of noise and allows one to estimate the energy gradient with a small amount of quantum circuits runs. As we pointed out before, each evaluation of the cost function requires the execution of only $3$ quantum circuits in our case. We perform $8192$ shots for each measurement of the cost function, finding it sufficient to successfully perform the optimization routine.
We then employ standard measurement error mitigation to compensate for readout errors \cite{Bravyi_2021}.
\begin{figure*} 
\begin{subfigure}{\columnwidth}
  \includegraphics[width=\columnwidth]{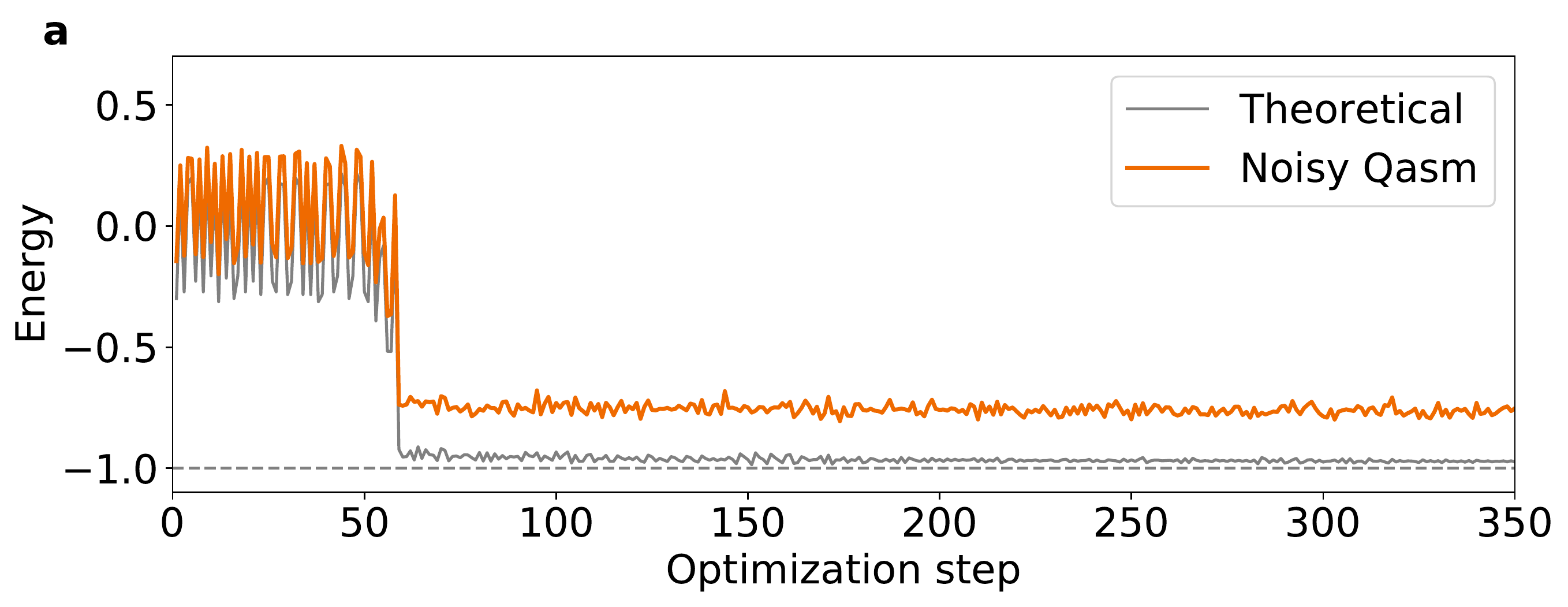}  
\end{subfigure}
\begin{subfigure}{\columnwidth}
  \includegraphics[width=\columnwidth]{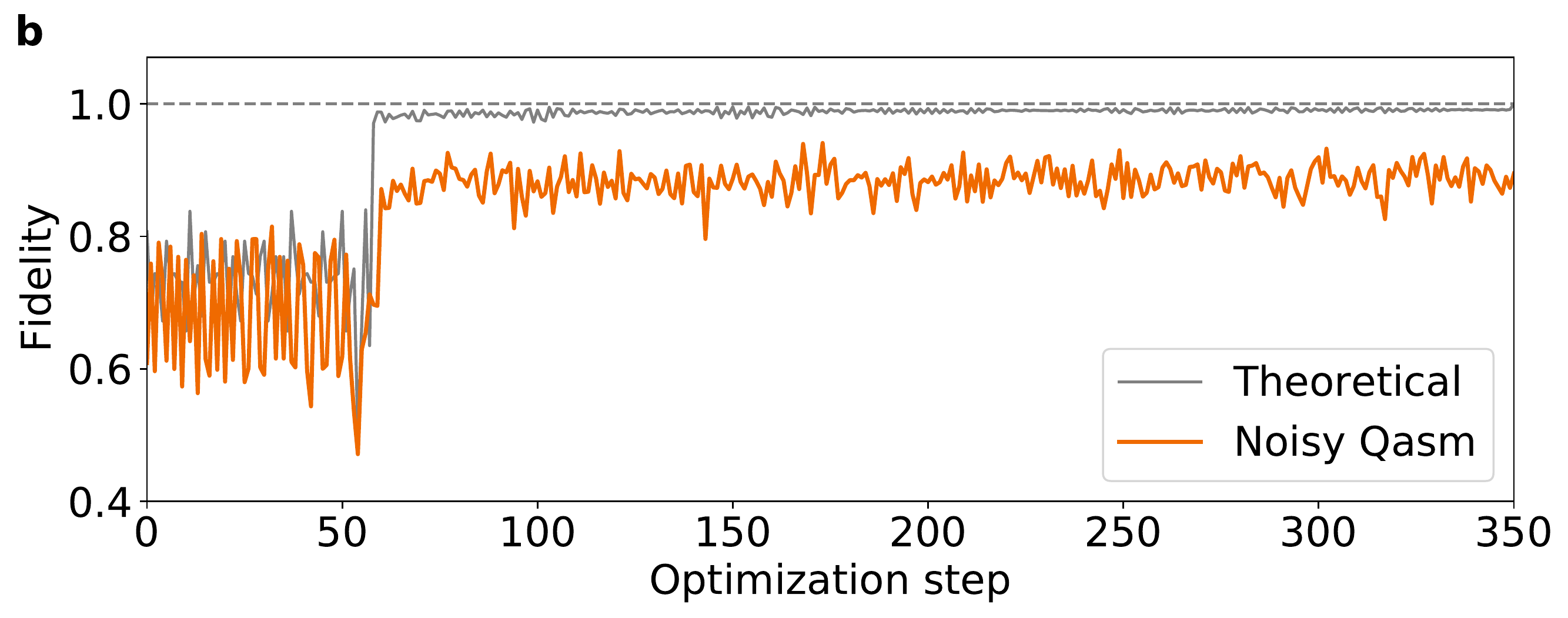}  
\end{subfigure}
\begin{subfigure}{\columnwidth}
  \includegraphics[width=\columnwidth]{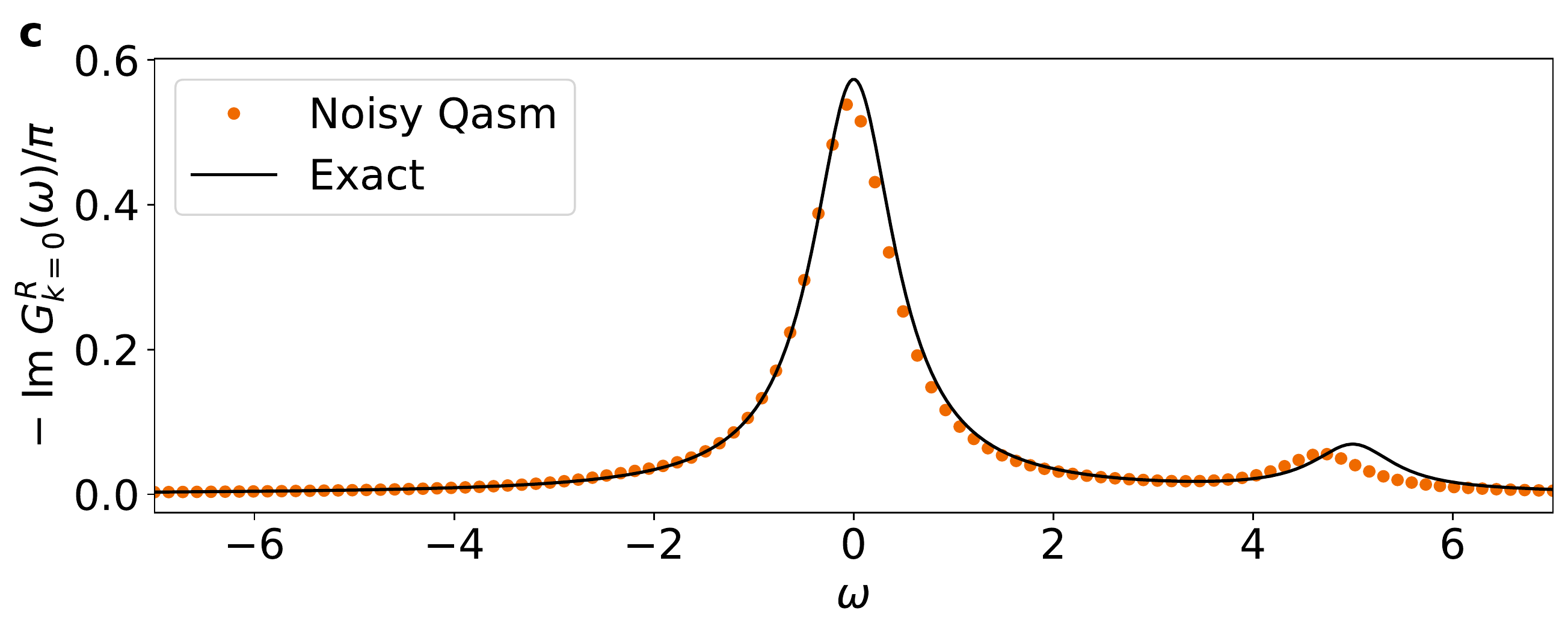}  
\end{subfigure}
\begin{subfigure}{\columnwidth}
  \includegraphics[width=\columnwidth]{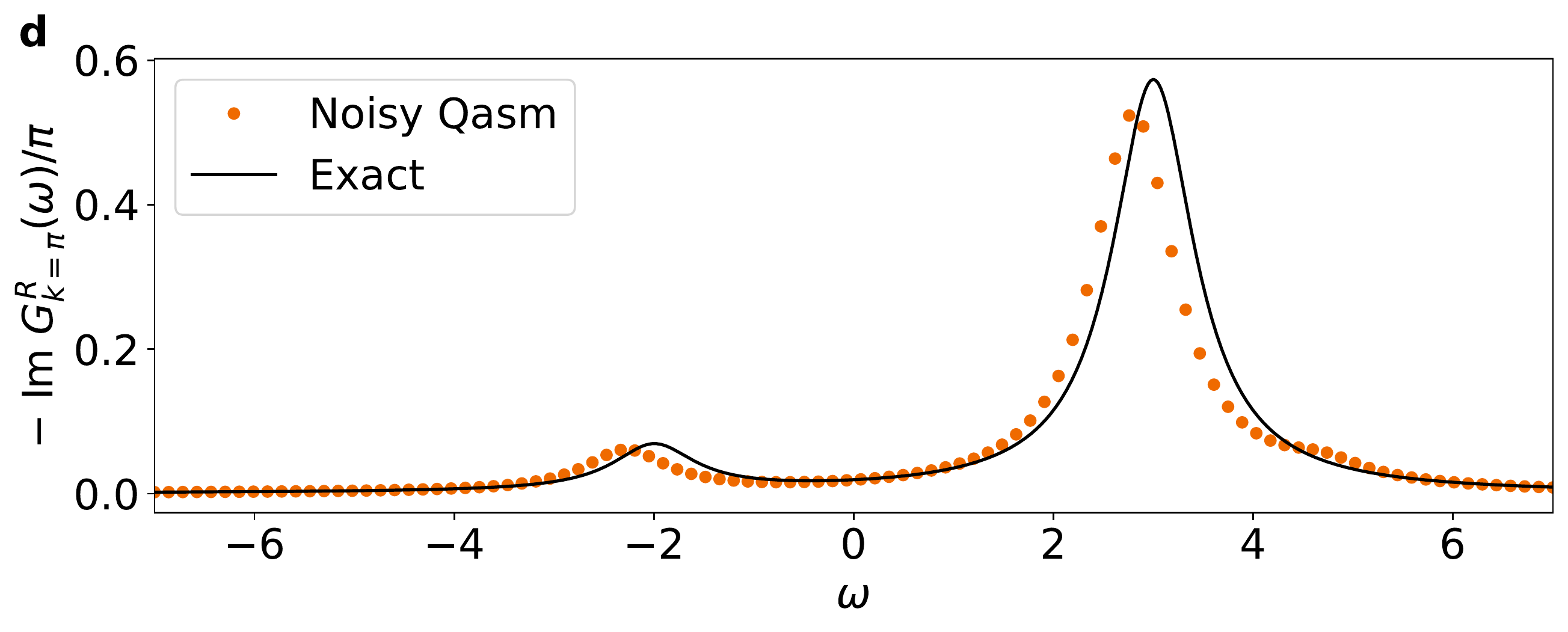}  
  \end{subfigure}
\caption{(\textbf{a}) Theoretical and noisy qasm estimate of the energy at each optimization step of the VQE procedure. The noise model is built to match the error rates of the \textit{ibm\_cairo} device. The dashed line is the exact ground state energy. (\textbf{b}) Theoretical and noisy qasm fidelity (obtained via QST) between the quantum state represented by the circuit at each time step and the exact ground state. The dashed line is the optimal fidelity. (\textbf{c})-(\textbf{d}) Noisy qasm simulation (and exact result) of the imaginary part of the retarded Green's function for $k=0$ (\textbf{c}) and $k=\pi$ (\textbf{d}) with $\eta=0.5$ via the qEOM algorithm.}
\end{figure*}
\begin{figure*}
\begin{subfigure}{\columnwidth}
  \includegraphics[width=\columnwidth]{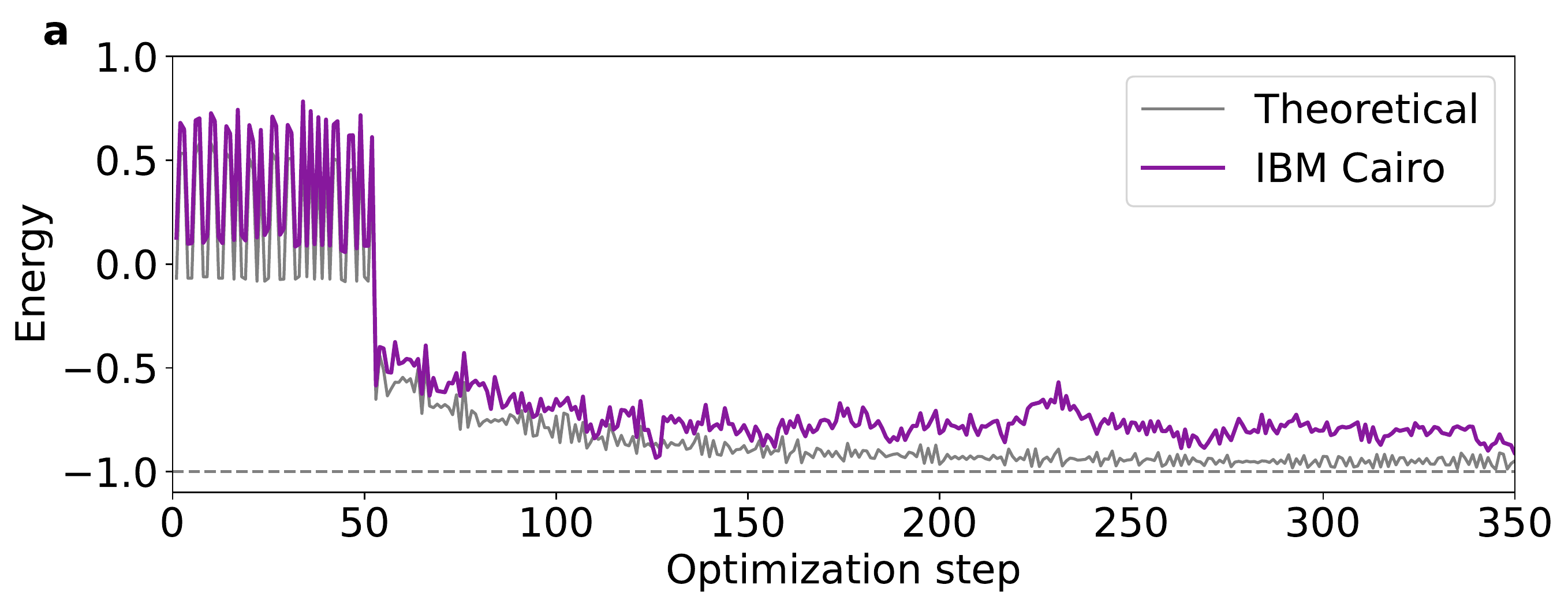}  
\end{subfigure}
\begin{subfigure}{\columnwidth}
  \includegraphics[width=\columnwidth]{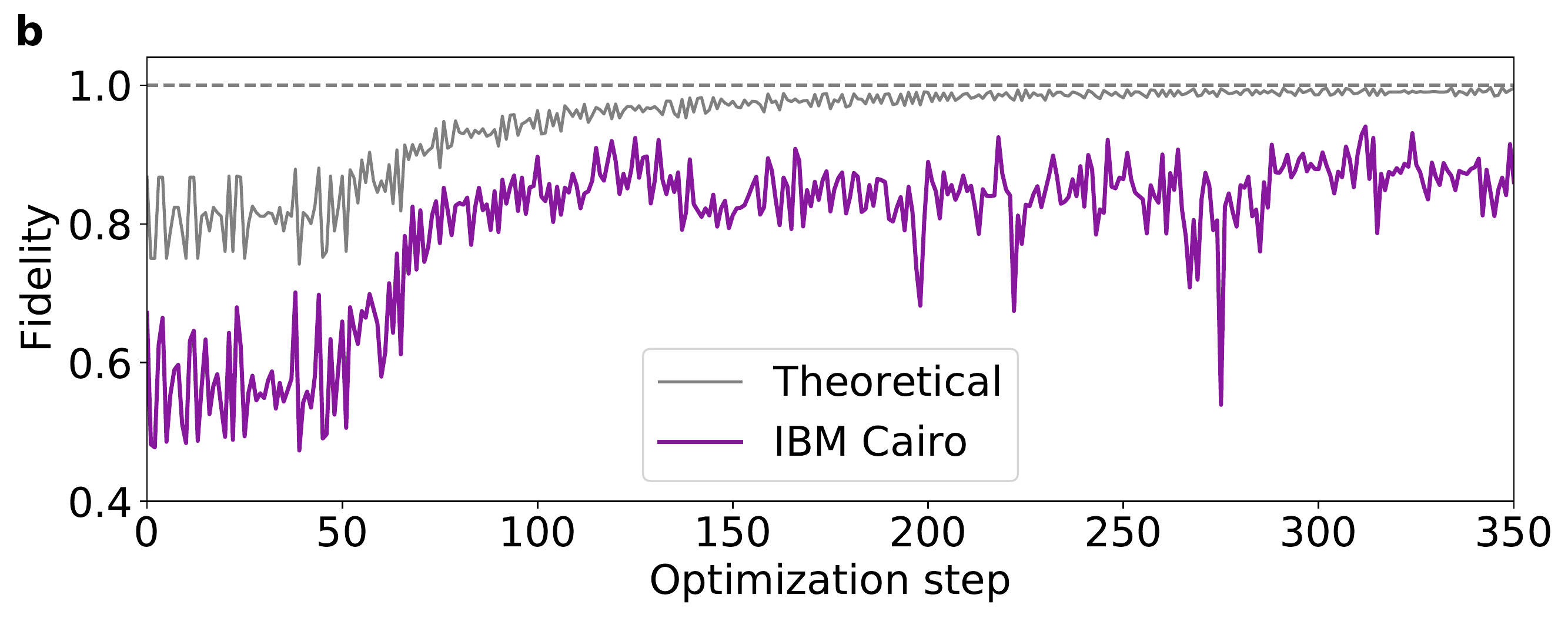}  
\end{subfigure}
\begin{subfigure}{\columnwidth}
  \includegraphics[width=\columnwidth]{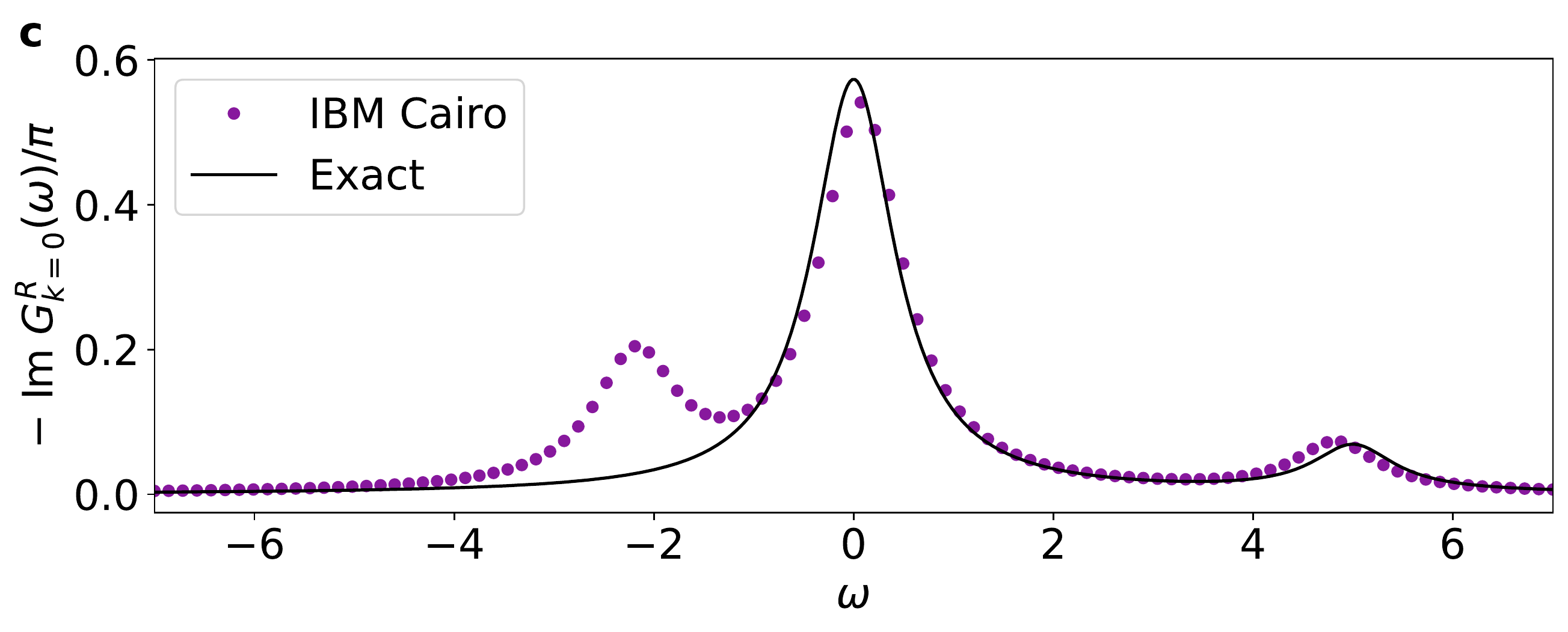}  
\end{subfigure}
\begin{subfigure}{\columnwidth}
  \includegraphics[width=\columnwidth]{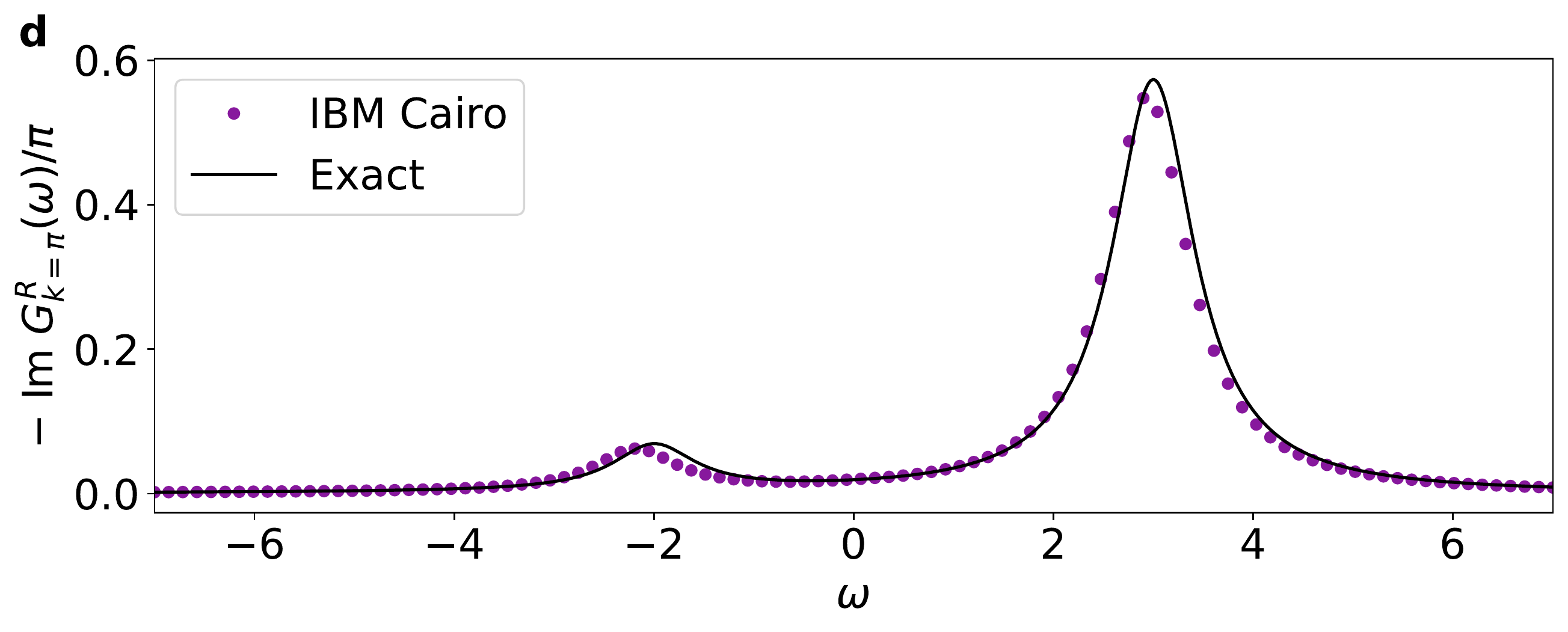}  
  \end{subfigure}
\caption{(\textbf{a}) Theoretical and experimental (\textit{ibm\_cairo}) estimate of the energy at each optimization step of the VQE procedure. The dashed line is the exact ground state energy. (\textbf{b}) Theoretical and experimental (\textit{ibm\_cairo}) fidelity (obtained via QST) between the quantum state represented by the circuit at each time step and the exact ground state. The dashed line is the optimal fidelity. (\textbf{c})-(\textbf{d}) \textit{ibm\_cairo} simulation (and exact result) of the imaginary part of the retarded Green's function for $k=0$ (\textbf{c}) and $k=\pi$ (\textbf{d}) with $\eta=0.5$ via the qEOM algorithm.}
\end{figure*}
In all simulations we use the ansatz in Fig.~1 and we iterate the VQE scheme until convergence. 
We plot the results of the VQE optimization procedure for different types of simulations in Fig.~3-6(a-b). We initialize the state randomly and we pick up the best VQE experiment in terms of the optimal energy reached over $10$ realizations. In Fig.~3-6(a), we plot the estimated energy of the parametrized state at each optimization step. For qasm, noisy qasm and hardware simulations we plot both the estimated energy (via projective measurements) and the theoretical energy, that is the energy computed via a statevector simulation of the ansatz circuit by setting all parameters at their actual values at the corresponding optimization step.

In Fig.~3-6(b), we plot the \textit{fidelity} between the approximate ground state and the true one at each optimization step $F(|\psi\rangle, | \phi \rangle) = | \langle \psi  | \phi \rangle |^2 \in [0,1]$.
\begin{table*}
\begin{center}
\begin{tabular}{ ccccc } 
\hline
 & Statevector & Qasm & Noisy qasm & \textit{ibm\_cairo} \\
\hline
Optimal energy & -1.0000  & -0.9945 (-0.9979)  & -0.78455 (-0.9995) & -0.8653 (-0.9871) \\ 
Optimal fidelity & 1.0000  & 0.9948 (0.9988)   & 0.8947(0.9998) & 0.8668 (0.9967) \\ 
\hline
\end{tabular}
\end{center}
\caption{Summary of the results of the VQE algorithm run with statevector, qasm, noisy qasm and \textit{ibm\_cairo} simulations. We show the results in the format \textit{measured (theoretical)} where \textit{measured} is the direct output from the experiment, while \textit{theoretical} is the value computed with statevector using the corresponding circuit parameters. All the simulations have been performed with 8192 shots and measurement error mitigation has been considered. The estimated fidelities have been obtained via QST.}
\end{table*}
The fidelity for the ground state is experimentally estimated by first extracting the coefficients of the quantum state via quantum state tomography (QST). Then the fidelity is computed analytically with respect to the exact ground state of the system. As for the energy estimate, for qasm, noisy qasm and hardware simulations we plot both the estimated fidelity (via QST) and the theoretical fidelity, obtained via the corresponding statevector simulation.
A summary of the results for the VQE is shown in Table~1. \\

Having obtained an estimate for the ground state of the system, we then implement the qEOM algorithm to reconstruct the charged excited states of the system and to compute the GF.
For the qEOM, we consider a set of $4$ excitation operators to retrieve the $4$ ($2$ particle and $2$ hole) excited states of the system. This is a novel extension to the originally proposed qEOM algorithm \cite{qeom_IBM}. The operators are chosen as follows: $\hat{E}_0 =  \hat{c}_{1,\downarrow}^{\dagger} \hat{c}_{1,\uparrow}^{\dagger} \hat{c}_{2,\uparrow}$, $\hat{E}_1 =  \hat{c}_{2,\downarrow}^{\dagger} \hat{c}_{1,\downarrow}^{\dagger} \hat{c}_{2,\downarrow}$, $\hat{E}_2 =  \hat{c}_{2,\downarrow}^{\dagger} \hat{c}_{1,\uparrow}^{\dagger} \hat{c}_{2,\uparrow}^{\dagger} \hat{c}_{1,\downarrow} \hat{c}_{2,\downarrow}$, $\hat{E}_3 =  \hat{c}_{2,\downarrow}^{\dagger}$.
We remark that other choices lead to equivalently good results. 
In our case, we get a 4 $\times$ 4 qEOM generalized eigenvalue problem, targeting simultaneously the two particle and two hole excited states. In particular, our selection of excitation operators targets only the particle states with $S_z = -1/2$ and the hole states with $S_z = 1/2$. This allows us to compute the GF selecting only the states with a specific spin symmetry, thereby reducing the size of the GEP to be solved.
The matrix elements entering Eq.~(13) are then estimated via projective measurements, after performing a JW mapping. 

We find the generalized eigenvalue problem to be quite sensitive to noise levels: indeed, small perturbations in the matrix elements may cause high changes in the qEOM eigenvalues that approximate the excitation energies. We alleviate this problem in two ways, first by increasing the number of measurements, hence improving the overall statistical precision, and then by measuring the excitation energies with a direct evaluation of the functional defined in Eq.~(11). In the second case, we find better estimates for the excitation energies with respect to the one obtained directly from solving the qEOM generalized eigenvalue problem. We further improve the results by averaging over $13$ realizations of the matrix elements estimation performed with $8192$ shots, giving an effective number of total shots of approximately $10^4$. We also notice that the general problem of minimizing the effect of noise in GEPs is well known in the literature~\cite{Hochstenbach_2019}, and we leave open for future studies the possibility of applying classical post-processing techniques to enhance the stability of the method.

Finally, we estimate the GF in the Lehmann representation by computing the spectroscopic elements in the form of Eq.~(16).
Since the excitation operators $\hat{O}_n$ can be expressed as a sum of 4 fermionic operators, we can decompose the whole operator $\hat{O}_n \hat{c}_{\alpha}^{\dagger}$ (or $\hat{O}_n \hat{O}_n^{\dagger}$) into a sum of Pauli strings and compute the spectroscopic elements by only applying single-qubit rotations at the end of the ansatz circuit (as for the operator averaging technique applied to VQE) and then performing projective measurements. 

In Fig.~3-6(c-d), we report the results for the GF computed in $k$ space, for $k=0,\pi$, that is considering the GF computed with $\hat{c}_{k,\sigma} = (\hat{c}_{1,\sigma}  +  e^{ik} \hat{c}_{2,\sigma})/\sqrt{2}$. In our case, we are actually evaluating the element $G_{k\uparrow,k\downarrow}$ of the GF, due to definition of the excitation operators. As shown in Fig.~3-6(c-d) we find a good agreement both between noisy simulations and the exact solution, and between the noisy qasm and the hardware simulations. This suggests that in this case the noise model does not severely underestimate the effect of noise in the real device. We also confirm the intuition that having a ground state which perfectly incorporates, despite the effect of noise, some conservation laws (i.e., spin and particle number) is crucial to ensure that the resulting excited states have the expected symmetries. However, while the eigenstates of the Hamiltonian corresponding to different energies should always be orthogonal, we find a considerable overlap between some of the evaluated excited states obtained via qEOM due to the approximate nature of our method. Since the Lehmann representation assumes an orthonormal basis of excited states, this has the effect of causing satellite peaks in the GF for noisy simulations, see Fig.~6(c-d). We leave deeper analysis and possible solution of this issue for future investigations.

\section{Conclusions}\label{sec:conclusions}
In this work, we propose a novel method for calculating the many-body one-particle GF compatible with near-term quantum devices. 
This method, in alignment with the work of Endo \textit{et al.}~in Ref.~\cite{PhysRevResearch.2.033281}, directly evaluates the transition amplitudes of the fermionic operators between the ground state and the charged excited states of the system. 
This is accomplished by an original extension of the qEOM algorithm introduced in Ref.~\cite{qeom_IBM}, which allows to compute the charged excited states of the Hamiltonian on a quantum computer, and relies on the preparation of the ground state obtained from the VQE.
The GF is evaluated in the frequency domain by classically combining the transition amplitudes and the excitation energies estimated via qEOM, with the use of the Lehmann representation of the GF. 

To test the validity of the present proposal, we applied this method to the two-site Fermi-Hubbard model and we performed both classical and quantum simulations of the algorithm, the latter on the \textit{ibm\_cairo} quantum processor.
We demonstrate that the proposed algorithm is able to correctly reproduce the target GF in such proof-of-principle experiments, although a rather large (compared to the usual VQE requirements) number of measurements is needed to obtain reliable estimates of the excitation energies, due to statistical and hardware noise. Although we expect this measurement overhead to become more prominent for larger systems when the standard form of our algorithm is considered, more effective strategies for observable reconstruction may be applied to meet such demand in an efficient way~\cite{Torlai_2020,huang2020predicting,hadfield2020measurements,garcia2021learning}, together with classical post-processing techniques aimed at enhancing the overall numerical stability and, possibly, the parallelization of the qEOM procedure.\\

All the steps involved in our proposal rely on estimating expectation values which can be efficiently evaluated on a quantum computer. While for the VQE the relevant observable to be estimated is the energy, in the qEOM algorithm the evaluation of the excited states relies on solving a GEP whose matrix elements cannot be efficiently obtained by classical means. At the same time, the dimension of the GEP problem scales only polynomially as the system size increases, thereby preserving the predicted quantum advantage. 

Since the qEOM result is sensitive to the accuracy of ground state preparation, substantial improvements over classical EOM methods are expected as soon as quantum computing methods will give access to those regimes where classical approximate ground state techniques perform poorly. Moreover, a considerable advantage of using qEOM to compute excited states compared to other quantum algorithms is that it does not require additional quantum resources (e.g., deeper circuits), with respect to the ground state VQE calculation. However, this must be balanced with an increase of the overall number of required measurements. 

It is important to stress that, under the proposed scheme, the quantum computation of the GF via the Lehmann representation is efficient only as long as a polynomial number of excited states is required for a reliable reconstruction of the relevant properties as the system size increases. Moreover, in practical implementations of the qEOM method described in this work one may also face a lack of orthogonality of the resulting excited states. While we have (efficiently) included the normalization factor in our simulations, adding methods to enforce orthogonality may deliver better results in the overall computation of the GF. 

Future research prospects include extending the proposed method to compute other types of GFs \cite{Fetter2003Quantum} and response functions, possibly at non-zero temperature, as proposed in Ref~\cite{PhysRevResearch.2.033281}, as well as scaling up the algorithm to larger quantum systems. The latter, other than being subject to general improvements of current quantum hardware, would require applying a full set of quantum error mitigation techniques~~\cite{temmeErrMit2017,Li_error_mit_prx_2017,kandala_error_2019,endo_practical_2018,suchsland2020algorithmic,gunther2021improving} and considering more efficient and application-specific ansätze~\cite{cadefermihubbard}. 
As already mentioned above, controlling the number of required measurements with effective observable reconstruction strategies~\cite{Torlai_2020,huang2020predicting,hadfield2020measurements,garcia2021learning} could also significantly extend the range of applicability of the qEOM approach.

\section*{Acknowledgments}

This research was supported by an E\textsuperscript{3} (EPFL Excellence in Engineering) program fellowship, the grants 200021-179312, 200021-179138 and NCCR MARVEL, funded by the Swiss National Science Foundation.

IBM, the IBM logo, and ibm.com are trademarks of International Business Machines Corp., registered in many jurisdictions worldwide. Other product and service names might be trademarks of IBM or other companies. The current list of IBM trademarks is available at \url{https://www.ibm.com/legal/copytrade}.

\bibliography{apssamp}

\end{document}